\documentclass[journal,article,accept,moreauthors,LaTex,dvi2pdf,10pt,a4paper,atoms]{mdpi}

\firstpage{1}
\articlenumber{x}
\doinum{10.3390/------}
\pubvolume{4}
\pubyear{2016}
\copyrightyear{2016}
\externaleditor{Academic Editors: A. Kumarakrishnan and Dallin S. Durfee}
\history{Received: 3 May 2016; Accepted: 16 June 2016; Published:}

 \theoremstyle{mdpi}
 \newcounter{thm}
 \newcounter{ex}
 \newcounter{re}
 \theoremstyle{mdpidefinition}

\usepackage{soul,microtype,upgreek}
    	
\Title{A Wigner Function Approach to Coherence in a Talbot-Lau Interferometer }

\Author{Eric Imhof $^{1, 2,}$*, James Stickney $^{1}$ and Matthew Squires $^{3}$}
\AuthorNames{Firstname Lastname, Firstname Lastname and Firstname Lastname}

\address{%
$^{1}$ \quad Space Dynamics Laboratory, Utah State University Research Foundation, North Logan, {UT 84341}, 
USA; {jim.stickney@gmail.com (J.S.)}
\\
$^{2}$ \quad {Kirtland Air Force Base, Albuquerque, NM 87117, USA}
\\
$^{3}$ \quad {U.S. Air Force Research Laboratory, Kirtland Air Force Base, Albuquerque, NM 87117, USA}; 
{Matthew.Squires@kirtland.af.mil (M.S.)}
}

\corres{Correspondence: eric.imhof@sdl.usu.edu; Tel.: +1-505-846-7260}

\abstract{Using a thermal gas, we model the signal of a trapped interferometer. This interferometer uses two short laser pulses, separated by time $T$, which act as a phase grating for the matter waves. Near time $2T$, there is an echo in the cloud's density due to the Talbot-Lau effect. Our model uses the Wigner function approach and includes a weak residual harmonic trap. The analysis shows that the residual potential limits the interferometer's visibility, shifts the echo time of the interferometer, and alters its time dependence. Loss of visibility can be mitigated by optimizing the initial trap frequency just before the interferometer cycle begins.}


\begin{document}




\section{Introduction}

Cold atom interferometry has been investigated for precision measurement
applications \cite{Barrett2011, Muntinga2013}, particularly  inertial navigation
\cite{Cahn1997, Cronin2009, Adams1993, Geiger2011}. Atom interferometers have
demonstrated orders of magnitude improvement in bias stability over
commercial navigation grade ring laser gyroscopes \cite{Durfee2006} and similar
gains are expected for accelerometers, gravimeters, magnetometers, and more.

Transitioning the technology to a real-world device has proven difficult. The
most sensitive atom interferometers use a 10-meter long apparatus \cite{Dickerson2013}.
These measurements rely on a Raman pulse technique which changes the internal
state of the interrogated atoms. Because of the difficulty in confining multiple states with a magnetic field,
atoms are allowed to propagate freely, necessitating a large system.

Single internal state splitting has allowed atoms to be trapped for the duration of the interferometer cycle, reducing the apparatus length to a few millimeters \cite{Wang2005}. Techniques for confined splitting include double-well potentials \cite{Schumm2006}, optical lattices \cite{Hilico2015}, and standing wave pulses \cite{Horikoshi2006, Burke2009}. However, these interferometers have used Bose-Einstein condensates, which require cooling stages that increase power consumption, decrease possible repetition rates, and lower atom numbers.

One single state technique has been shown to work at thermal (\emph{i.e.}, non-condensed) temperatures~\cite{Wu2005,Xiong2013,Mok2013}. These interferometers, in the
``Talbot-Lau'' configuration, confine the atomic sample in two directions and
allow free propagation in the third. In an ideal situation, the potential
along the third direction would vanish. However, due to the finite size of the
device and uncontrollable external fields, there is residual potential along
the waveguide.

\scalebox{.95}[.95]{Unfortunately, the residual potential and other field imperfections reduce coherence times \cite{Burke2009, Wu2007, Su2010}.} Recent
research has demonstrated a high degree of control over the residual field \cite{Stickney2014}. Here, we analyze the effect of a controlled residual
potential in a Talbot-Lau interferometer with a gas of cold, thermal atoms
using a Wigner function approach.

\section{Interferometer Operation}

To prepare the atomic gas for the interferometer cycle, a laser cooled sample
is loaded into a magnetic trap with frequencies $\omega_i^{(e)}$, where $i
= (x,y,z)$. The collision rate is directly proportional to the geometric
average of these trap frequencies $\bar {\omega}^{(e)} = (\omega^{(e)}_x
\omega^{(e)}_y \omega^{(e)}_z)^{1/3}$, so $\bar {\omega}^{(e)}$ should be made
as large as possible to maximize the efficiency of the evaporative cooling. In
typical atom chip experiments, the gas is evaporatively cooled in a trap with
frequency $\bar{\omega}^{(e)} \sim
2 \pi \times 200~\mbox{Hz}$.

Once the atoms are cooled to a temperature on the order of $\mathcal{T} \sim
10~\upmu\mbox{K}$, the potential is adiabatically transformed into a trap that
tightly confines the atoms in the radial direction, with frequencies $\omega_y
= \omega_z = \omega_\perp \sim 2 \pi \times 200~\mbox{Hz}$; and in the axial
direction, with frequency $\omega_x = \omega_0$. Just before the interferometer
cycle starts, the potential is non-adiabatically transformed into a waveguide
potential, while holding the radial trap frequency constant to reduce the
effects of transverse excitations. In a realistic device, there remains
a residual potential along the waveguide with frequency $\omega$.

\begin{figure}[H]
\centering
\includegraphics[scale=.5]{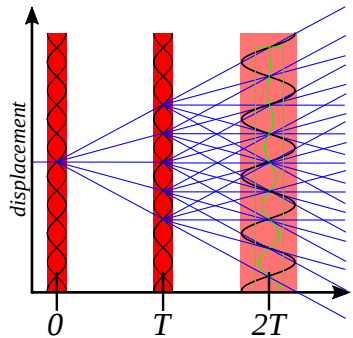}
\caption{The schematic of a Talbot-Lau interferometer.  An atomic cloud is split in space (vertical axis) by a laser pulse at time $t = 0$.  The resulting diffracted orders separate, and are further diffracted at $t = T$.  At the recombination time $t = 2T$, the various orders overlap, allowing a probe laser to produce a back scattered signal from the periodic atomic distribution.  We only show two diffraction orders because for typical laser pulses, higher orders are suppressed.
\label{fig:TLI} }
\end{figure}

Once the atoms are loaded in the waveguide, the interferometer cycle begins.
In this analysis, we considered the case of the trapped atom Talbo t-Lau interferometer
schematically shown in Figure~\ref{fig:TLI}. The figure traces the
different paths that an initially stationary atom could experience when moving
through the device. Time moves from left to right, and the displacement of the
atom along the waveguide is shown in the vertical direction.

At time $t=0$, the atomic cloud is illuminated with a short, standing
wave laser pulse that acts as a  diffraction grating. The pulse is
sufficiently short that it is in the Kapitza-Dirac regime, \emph{i.e.},\ the atoms do
not move for the duration of the laser pulse. The pulse splits the wave
function for each atom into several momentum states separated by the
two photon recoil momentum $\delta P = 2 \hbar k_l$, where $k_l$ is the wave
number of the laser beams.

After the laser pulse, the atomic cloud propagates in the waveguide for a time $T$,
at which point it is illuminated with a second laser pulse. The paths of the
different momentum states are shown as blue lines between $0$ and $T$.
Ideally, the momentum of each mode should be constant in time. However, the
residual curvature along the waveguide will cause the paths to become curved (not shown in the figure), giving rise to decoherence.

For simplicity, it is assumed that the laser pulse at time $T$ has the same strength and
affects the atomic wave function in the same manner. Each of the momentum states
that were populated after the first laser pulse are split into several
modes. After the second laser pulse, the number of possible paths
increases dramatically. However, near time $2T$, the different
paths come together to form a density modulation that has the same period as the
standing wave.

An extraordinary feature of a Talbot-Lau interferometer is that the location
of the density echo is independent of the initial velocity of the atom. For example,
if the initial atom in Figure~\ref{fig:TLI} had some momentum, each of
the diffracted orders would gain this additional momentum. After tracing out
all possible paths, it is easy to show that the density modulation appears in
exactly the same location as for the initially stationary atom.
As a result, the density echo is still visible even when the initial atomic
gas is relatively hot.

In the absence of external forces, the density echo will have the same relative
phase as the standing wave laser pulse. However, if there is a force on the
cloud, the echo will move in response to the force. By detecting the shift in
the echo, it is possible to measure the force on the cloud.

This phase shift can be measured by reflecting a traveling wave off the density
modulation. Due to the Bragg effect, there will be a strong backscattered signal
for the duration of the echo. By~heterodyning the back-reflected light with a
reference beam, the phase of the density echo can be~determined.

In this paper, we present a theoretical model of a trapped Talbot-Lau
interferometer that includes the decoherence due to the residual potential
curvature. We use the Wigner function approach to model the dynamics of
a thermal gas, which can be extended to include more complex laser
pulse sequences \cite{Su2010}. For brevity, only the simple case of
a two-pulse interferometer is discussed. Our model predicts the
amplitude of backscattered light for an arbitrary initial Wigner function and is then specialized
to the case of an initial thermal distribution.
Decoherence due to finite temperature and initial axial trap frequency are
discussed. Finally the model is used to determine the ideal axial frequency
for a given initial phase space density and residual potential.

\section{The Model}

Following the prescription of \cite{Stickney2014}, we assume that the potential
is separable, \emph{i.e.}, $V(\pmb r) = V(x) + V_\perp(r_\perp)$, and the $k$-vectors of
the laser beams point in the $x$-direction. Collisions are neglected as we
have previously analyzed the effects of collisions in a similar interferometer
and do not expect atom-atom collisions to have a significant impact on
the results \cite{Stickney2009}. We also ignore the mean field interaction, as it is mainly relevant for strongly interacting condensates, which we do not consider here.  Inclusion of these terms may be possible, but are omitted to keep the discussion concise.  The Hamiltonian that governs the axial
dynamics of the interferometer is one-dimensional and can be written as
\begin{equation}
    H = \frac{P^2}{2M } + \frac{1}{2} M \beta X^2 + \hbar \Omega \cos(2 k_l X),
    \label{eqn:dimensionalHamiltonian}
\end{equation}
where $X$ and $P$ are the canonical operators with commutation relation $[X, P]
= i \hbar$, $k_l$ is the wave number of the laser, $M$ is the atomic mass, and
$\beta$ is the curvature of the residual potential. The parameter $\Omega$ is
the frequency of the AC-stark shift due to the standing wave laser pulse, which
depends on the intensity and detuning of the beam and is, in general,
a function of time.

The Hamiltonian can be recast in the dimensionless form
\begin{equation}
    H' = \frac{P'^2}{2} + \frac{1}{2} \beta' X'^2 + \Omega' \cos X',
    \label{eqn:dimensionlessHamiltonian}
\end{equation}
where $P' = P/P_0$, $X' = X/X_0$, and $t' = t/t_0$ where $P_0 = 2 \hbar k_l$,
$X_0 = 1/2 k_l$, and $t_0 = M / 4 \hbar k_l^2$. The other parameters in
Equation~(\ref{eqn:dimensionalHamiltonian}) become $\beta' = \beta t_0^2$ and,
$\Omega' = \Omega t_0$. The other important dimensionless parameter is the
cloud temperature $\mathcal{T}' = \mathcal{T} / \mathcal{T}_0$, where
$\mathcal T_0 = 4 \hbar^2 k_l^2 / M k_B$, where $k_B$ is the Boltzmann constant.
For $^{87}\mbox{Rb}$ where the standing wave laser is near the D2 transition,
$t_0 = 5.3~\upmu\mbox{s}$, and $\mathcal{T}_0 = 1.4~\upmu\mbox{K}$. For the rest of this paper, primes will be dropped for clarity, and unless otherwise stated, all introduced variables will be dimensionless.

Since the interferometer uses an incoherent gas, the state of the system cannot
be written as a wave function. Instead, the system is described by the density
operator $\rho$. The equation of motion for the density operator, in
dimensionless form, is
\begin{equation}
    i \dot \rho = [H, \rho],
    \label{EQM_Rho}
\end{equation}
where the dot denotes the time derivative and the brackets are the usual commutation
operator.
The density operator can be recast in terms of the Wigner function, which
is defined as
\begin{equation}
    f(x,p) = \frac{1}{\pi} \int d\xi \langle x + \xi|\rho|x - \xi\rangle
            e^{-2i p \xi}
    \label{WignerFunction}
\end{equation}
where $|x\rangle$ are the eigenvectors of the coordinate operator, \emph{i.e.}, $X
|x\rangle = x |x\rangle$. The Wigner function $f(x,p)$ can be interpreted as
the probability density, however for non-classical states the Wigner function
may be negative. As a result, $\int dx f = \mathcal{P}(p)$
is the momentum density of the cloud and $\int dp f = \rho(x)$ is the spatial
density. Even when the Wigner function is negative, the densities,
$\mathcal{P}$ and $\rho$  are always positive.

It is worth noting that the Wigner approach works for pure states as well.  In this case, it is defined~as
\begin{equation}
f_{pure}(x,p) = \frac{1}{\pi} \int d\xi \psi^*(x+\xi)\psi(x-\xi)e^{-2ip\xi}.
\label{eq: pure}
\end{equation}
We will find that the results of the incoherent process are easily extended to include the results of a pure state (BEC) interferometer.

Substituting Equation~(\ref{WignerFunction}) into Equations~(\ref{eqn:dimensionlessHamiltonian}) and
(\ref{EQM_Rho})
it can be shown that the equation of motion for the Wigner
function is
\begin{equation}
    \left( \frac{\partial}{\partial t} + p \frac{\partial}{\partial x}
    - \beta \frac{\partial}{\partial p}\right) f(x,p, t) =
    \Omega \sin x \left[ f\left(x, p - \frac{1}{2} \right)
    - f\left( x, p + \frac{1}{2} \right) \right],
    \label{eqn:EQM_WignerFunction}
\end{equation}
where the left side of the equation describes the motion of the distribution
in the potential while the right side describes the interaction with the
standing wave laser field.

Since the duration of the laser pulses $\tau_p$
is much shorter than the interferometer time $T$ ($T \gg \tau_p$), the evolution
of the distribution can be separated into relatively slow dynamics when the
distribution is not being illuminated and fast dynamics when it is.
Additionally, since each laser pulse is short $\tau_p
\gg 1/\omega_0$ and strong $\Omega \gg \omega_0$, the pulses are in the
Kapitza-Dirac regime, which occurs in the Raman-Nath limit. As a result,
the coordinate and momentum derivatives in
Equation~(\ref{eqn:EQM_WignerFunction}) may be neglected during the pulse. \enlargethispage{.5cm}

The dynamics of the distribution for the periods when the laser is off, $\Omega
= 0$, are such that each part of phase space evolves classically.
For simplicity, it is useful to write the classical equations of motion in the form
\begin{equation}
    \dot {\pmb x} = M \pmb x
    \label{eqn:NSL}
\end{equation}
where $\pmb x = (x, p)$ is the coordinate-momentum vector, and the matrix $M$
is
\begin{equation}
    M = \left(
    \begin{array}{cc}
        0 & 1 \\
        -\beta & 0
    \end{array}
        \right).
    \label{}
\end{equation}
The solution to Equation~(\ref{eqn:NSL}) can be written as $\pmb x(t) = U_t \pmb
x(0)$, where $U_t = \exp(M t)$. By direct substitution it can be shown that in between
the laser pulses the distribution evolves as
\begin{equation}
    f_f( \pmb{x} ) = f_i(U_{-t} \pmb x).
    \label{eq:freeProp}
\end{equation}

The laser pulses are more involved and fundamentally quantum in nature (\emph{i.e.},\
resulting in negative Wigner distributions). The effect of the laser pulse is
to transform an initial Wigner distribution $f_i$ into a final distribution
$f_f$ according to

\begin{equation}
     f_f(\Omega \neq 0) = \sum_{nk = -\infty}^\infty (-i)^n
     J_k(\Xi)J_{n+k}(\Xi)e^{i(n+2k)x}
        f_i\left( x, p - \frac{n}{2} \right)
    \label{eq:LaserPulse}
\end{equation}
for the pulse area, $\Xi = \int d\tau \Omega(\tau)$, where the
functions $J_n$ are the Bessel functions of the first kind.
In terms of $\pmb x$, Equation~(\ref{eq:LaserPulse}) can be written
in the more compact form
\begin{equation}
    f_f(\Omega \neq 0) = \sum_{nk} \alpha_{nk}
    e^{i \pmb{g}_{nk} \cdot \pmb{x}} f_i(\pmb{x} - \textbf{N}_n)
\label{eq:LaserPulse2}
\end{equation}
where
$\pmb {g_{nk}} =  (n+2k, 0)$, $\pmb N_n = (0, n/2)$,
and $\alpha_{nk} = (-1)^n J_k J_{n+k}$.

The interferometer sequence is characterized by four unique operations
separated in time. The first laser pulse at $t = 0$ operates on an initial
Wigner distribution $f_0$ and transforms it to $f_1$, ($f_0 \rightarrow f_1$).
There is then a propagation period from $t = 0$ to $T$, over which the
distribution transforms $f_1 \rightarrow f_2$. The second laser pulse at $t=T$
transforms $f_2 \rightarrow f_3$. Lastly, another propagation to $t
= 2T+\tau$ transforms the distribution to its final form $f_3 \rightarrow f_4$.

Near the end of the interferometer cycle, the cloud is illuminated with
a short traveling wave laser pulse of duration $\tau_0$, where $\tau_0 \ll T$.
To determine the time dependence of the back-scattered light, the Wigner function
must be found for times near the echo time, \emph{i.e.}, $t = 2 T + \tau$.
By direct substitution into Equations~(\ref{eq:freeProp}) and  (\ref{eq:LaserPulse2})
for the interferometer cycle discussed in Figure~\ref{fig:TLI}, the Wigner function
near the echo time is
\begin{eqnarray}
    f_{4} &=&
\sum_{mlnk} \alpha_{ml} \alpha_{nk} \nonumber \\
   &\times&
    \exp\left[
        i (\pmb{g}_{ml} \cdot U_{T}
        + \pmb{g}_{nk} )\cdot U_{-2 T - \tau} \cdot \pmb{x}
        -  i \pmb{g}_{nk} \cdot U_{-T} \cdot \pmb{N}_m
        \right]
\nonumber \\
&\times& f_0(U_{-2T-\tau} \cdot \pmb{x} - U_{-T}\cdot \pmb{N}_m - \pmb{N}_n).
\label{eqn:finalWigner}
\end{eqnarray}
According to \cite{Wu2007}, the amplitude of the back-scattered light
is proportional to
\begin{equation}
    S = \int d^2x e^{i \pmb g_{01}\cdot \pmb x} f_{4} (\pmb{x})
    \label{eqn:sig}
\end{equation}
For the rest of the paper, the quantity $S$ will be referred to as the
signal of the interferometer.
Changing the integration variable from $\pmb x$ to $\pmb y$, where
\begin{equation}
\pmb{y} = U_{-2T-\tau}\pmb{x} - U_{-T}\textbf{N}_m - \textbf{N}_n,
\end{equation}
the signal can be written as
\begin{equation}
    S = \sum_{mlnk} \alpha_{ml}\alpha_{nk}
    e^{i \Theta_{mlnk}} \int d^2 y e^{i \pmb \Delta_{mlnk} \cdot \pmb y},
    \label{}
\end{equation}
where
\begin{equation}
    \pmb \Delta_{mlnk} = \pmb g_{ml} \cdot U_T + \pmb g_{nk} + \pmb g_{10} \cdot U_{2T + \tau},
    \label{eq:DELTA}
\end{equation}
and
\begin{equation}
    \Theta_{mlnk} = \pmb \Delta_{mlnk} \cdot \left( U_{-T} \cdot \pmb N_m
                             + \pmb N_n
                             \right)
                             - \pmb g_{nk} \cdot U_{-T} \cdot \pmb N_m.
    \label{}
\end{equation}

In what follows below, it will be assumed that both the echo duration
is small as compared to the interferometer time $\tau \ll T$, and the
residual trap curvature is  $\beta \ll 1/T^{2}$. When these inequalities
are fulfilled, only the linear contributions in both $\tau$ and $\beta$
are retained.
In this limit, the time propagation operator for small values of $\beta$ is
$U_T \approx U_T^{(0)} + \beta U_T^{(1)}$, where
$U_t^{(0)} = \left(\begin{smallmatrix}  1 & t \\ 0 & 1  \end{smallmatrix}\right)$,
and
$U_t^{(1)} = \left(\begin{smallmatrix}  t^2/2 & t^3/6 \\ t & t^2/2 \end{smallmatrix} \right)$,
and for small values of time $\tau$, $U_\tau = 1 + M^{(1)}\tau$,
where
$M^{(1)} = \left(\begin{smallmatrix}  0 & 1 \\ 0 & 0  \end{smallmatrix}\right)$.

Equation~(\ref{eq:DELTA}) can now be written as
\begin{equation}
    \pmb \Delta_{mlnk} =
    \pmb \Delta_{mlnk}^{(0)}
    + \beta\left( \pmb g_{ml} \cdot U^{(1)}_T + \pmb g_{10} \cdot U^{(1)}_{2T} \right)
    + \tau \left( \pmb g_{10} U_{2T}^{(0)} M^{(1)} \right),
    \label{deltaApprox}
\end{equation}
where $\Delta^{(0)}$ is given by Equation~(\ref{eq:DELTA}) where $\beta \rightarrow
0$ and $\tau \rightarrow 0$. In the limit where the distribution is slowly
varying, the elements of the sum in Equation~(\ref{eqn:sig}) are vanishingly small
unless $\Delta^{(0)} = 0$. This~implies that $\pmb g_{ml} = -2 \pmb g_{10}$ and
$\pmb g_{nk} = \pmb g_{10}$. Using the definition of $\pmb g$, these relations can
be written as $k = (1-n)/2$ and $l=-(2+m)/2$. In addition, only the terms where
$n$, ($m$) are even (odd) contribute to the signal.
Equation~(\ref{deltaApprox}) becomes independent of the indices $m,l,n,k$.

Substituting the explicit matrix representations for $\Delta$ and $\Theta$, the
interferometer signal is given by
\begin{equation}
    S = A\int dudv ~ \mbox{exp}\left[ -i\beta T^2u
               + i \tau' v \right]f_0(u,v),
    \label{signal}
\end{equation}
where $\tau' = \tau - \beta T^3$ and $u,v$ are the components of the vector $\pmb{y}$. The parameter $A$ in Equation~(\ref{signal}) is the amplitude of the signal and can be
expressed as the sum
\begin{equation}
    A = \sum_{n, {even}} \sum_{m, {odd}} \frac{\gamma_{nm}}{2i}
    \exp\left[ i\left( \frac{m T}{2}
                    + \frac{m+n}{2}\tau'
                     + \frac{5m }{12} \beta T^3 \right) \right],
    \label{A}
\end{equation}
where
\begin{equation}
    \gamma_{nm} = 2 (-1)^{(n-1)/2+m/2} J_{(1-n)/2} J_{(1+n)/2} J_{-(2+m)/2}J_{-(2-m)/2}
    \label{}
\end{equation}
determines proportion of the atoms scattered into each mode.

Equation~(\ref{signal}) is the primary result of this analysis, and will be used for the case of a thermal atomic cloud in Section \ref{discussion}.

\section{Discussion}
\label{discussion}

By taking the limit where $\Xi \ll 1$, only the lowest order contributions to
Equation~(\ref{A}) need to be retained. If we keep $n= \pm 1$ and $m = 0, \pm 2$ and
use the limiting values of $J_n$ for the small argument,
$\gamma_{10} \approx - 2 \gamma_{12} \approx \Xi / 4$, then
\begin{equation}
     A =  \sin\left( \frac{1}{2} \tau' \right)
     \frac{\Xi^3}{4}
     \left[ 1 +
         \cos\left( T + \frac{5 \beta T^3}{6} + \tau ' \right)
     \right].
     \label{Aapprox}
 \end{equation}

 Assuming that  the initial distribution is a thermal cloud of
 temperature $\mathcal{T}$ that is in
equilibrium with the trap with frequency $\omega_0$, the distribution $f_0$ becomes
\begin{equation}
    f_0 = \frac{\omega_0}{2 \pi \mathcal{T}}
    \exp\left(- \frac{p^2}{2 \mathcal{T}}
    - \frac{\omega_0^2 x^2}{2 \mathcal{T}} \right).
    \label{ThermalDist}
\end{equation}

By comparison, the initial distribution of a condensate would be well approximated by the ground state of a harmonic oscillator.  Using Equation (\ref{eq: pure}), the pure state Wigner function is equivalent to Equation (\ref{ThermalDist}) when $\mathcal{T} = \omega_0/2$.  During the transition from an incoherent thermal gas to a pure BEC, the distribution is a sum of $f_0$ and $f_{pure}$, weighted by the number of atoms in and out of the ground state, where $N_0/N = 1- (\mathcal{T}/\mathcal{T}_c)^3$ and $\mathcal{T}_c = \bar{\omega}_0 (N/\zeta(3))^{1/3}$.  $N_0/N$ is the ratio of condensed atoms to the total, and $\zeta$ is the Riemann zeta function. This combined distribution can be used with Equation~(\ref{signal}) to find the expected signal.

Returning focus to the incoherent thermal gas, substituting Equations~(\ref{Aapprox}) and (\ref{ThermalDist})
into Equation~(\ref{signal}) and performing the integral yields
\begin{equation}
    S = A
    \exp\left[-\frac{\mathcal{T}}{2} \left(\frac{\beta T^2}{\omega_0}\right)^2
    -\frac{\mathcal{T}}{2} \tau'^2 \right].
    \label{SS}
\end{equation}

To quantify the signal visibility, we define the echo strength as $\mathcal{I} = \Xi^{6} \int d\tau' S^2$, which
is proportional to the total number of photons (electromagnetic energy) of the backscattered
light during the read-out pulse. In the limit where $\mathcal{T} \gg 1$,
Equation~(\ref{SS}) can be integrated, yielding
\begin{equation}
    \mathcal{I} = \frac{\pi^{1/2}\mathcal{A}^2} {32 \mathcal{T}^{3/2}}
    \exp\left[-\mathcal{T} \left(\frac{\beta T^2}{\omega_0} \right)^2 \right],
    \label{I0}
\end{equation}
where
\begin{equation}
    \mathcal{A} = \frac{1}{4} \left[ 1 + \cos\left( T + \frac{5}{6}\beta T^3 \right) \right].
    \label{<++>}
\end{equation}
Equation~(\ref{I0}) diverges in the limit $\mathcal{T} \rightarrow 0$, which is
clearly an unphysical result. However the numerical integration of
Equation~(\ref{SS}) remains finite.

\begin{figure}[H]
\centering
\includegraphics[scale=.7]{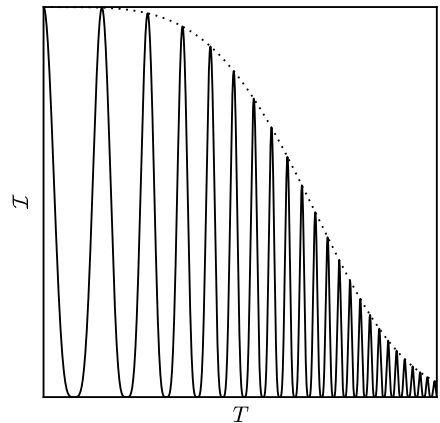}
\caption{
    A schematic of the echo signal strength, $\mathcal{I}$, as a function of interferometer time, $T$, for an interferometer in a positive residual trapping potential, \emph{i.e.}, $\beta>0$.  The signal strength is proportional to the total number of backscattered photons during the readout laser pulse.  $\mathcal{I}$ is periodic with an increasing frequency within an envelope defined by the dotted curve.
}
\label{fig:IofT2}
\end{figure}

\begin{figure}[H]
\centering
\includegraphics[scale=.7]{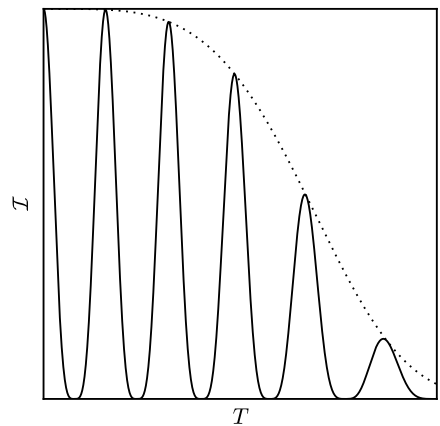}
\caption{
    A schematic of the echo signal strength, $\mathcal{I}$, as a function of interferometer time, $T$, for an interferometer in a negative residual trapping potential, \emph{i.e.}, $\beta<0$.  Like the positive potential case, the signal strength is periodic and contained within a decaying envelope.  However, the negative potential causes a decreasing frequency.  Both positive and negative potentials have the same envelope.
}
\label{fig:IofT2m}
\end{figure}

Note that $\mathcal{I}$ is an oscillating function, and is well known in the
$\beta = 0$ case \cite{Cahn1997}. Figure~\ref{fig:IofT2} shows a schematic of
Equation~(\ref{I0}) as a function of interferometer time $T$ for $\beta >0$. The
dotted line is the envelope of the echo strength. Figure~\ref{fig:IofT2m}
shows a schematic of Equation~(\ref{I0}) as a function of interferometer time $T$
for $\beta <0$.

The oscillation frequency increases when $\beta > 0$ and decreases when $\beta
< 0$, and there is a maxima when $T + 5 \beta T^3 / 6 = 2 \pi n$. These
oscillations depend only on the values of $\beta$ and $T$. In a typical
experiment, the oscillation frequency is much larger than depicted in
Figure~\ref{fig:IofT2} or Figure~\ref{fig:IofT2m}. For the remainder of the paper,
it will be assumed that the interferometer time is tuned to be at the peak of
an oscillation, which will be referred to as $\mathcal{I}_m$.

In order to maximize signal strength, it is also useful to release the atomic
sample into the waveguide from the correct initial trap. Typically, the atomic
gas is evaporatively cooled to a temperature $\mathcal{T}^{(e)}$ in a trap with
frequency $\omega^{(e)}$. After cooling, the trap frequencies are
adiabatically changed to a trap with frequency $\omega_0$ and then released
into a waveguide with residual potential curvature $\beta$. During the
adiabatic transformation, the phase space density is constant. This condition
implies $D = \mathcal{T}^3/ \omega$ is held constant, assuming the radial trap
frequencies $\omega_\perp$ are unchanged. Then Equation~(\ref{I0}) can be recast as
\begin{equation}
    \mathcal{I}_m = \frac{\pi^{1/2}}{32 (D^{(e)})^2 \omega_0^2}
    \exp\left[ - \frac{(D^{(e)})^{1/3}}{\omega_0^{5/3}} (\beta T^2)^2 \right],
    \label{I000}
\end{equation}
where $D^{(e)} = (\mathcal{T}^{(e)})^3 / \omega^{(e)}$ is proportional to
the phase space density at the end of the evaporation.

\begin{figure}[H]
\centering
\includegraphics[scale=.7]{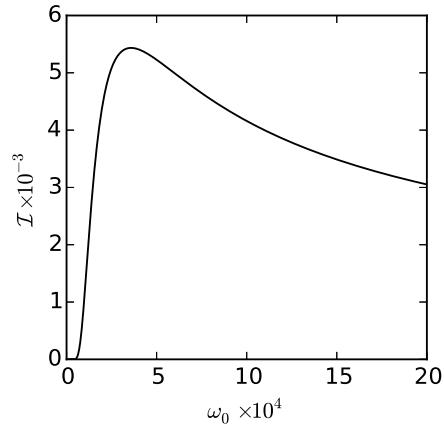}
\caption{
    The signal strength as a function of injection trap frequency, $\omega_0$.  After evaporation in a trap with frequency $\omega^{(e)}$, the trap potential is adiabatically transformed to $\omega_0$ before the interferometer cycle begins.  At the start of the cycle, the trap is snapped to $\omega = \sqrt{\beta}$, where it stays. The signal strength peaks at for a non-zero injection frequency $\omega_0$. For this case, $\beta T^2 = 10^{-4}$, and $D^{(e)} = 10^6 / 2 \pi$.
}
\label{fig:IoOmega0}
\end{figure}

For this analysis, assume the cloud is evaporatively cooled in a trap with frequency \mbox{$\omega^{(e)}
= 2\pi \times 10^{-4}$} and to a temperature $\mathcal{T} = 10$. For
$^{87}\mbox{Rb}$, these parameters correspond to a gas cooled in a trap with
a frequency of $20~\mbox{Hz}$ to a temperature of $14~\upmu\mbox{K}$. The phase
space density is proportional to $D^{(e)} = 10^{-6}/2 \pi$. Figure~\ref{fig:IoOmega0}
shows the echo strength, Equation~(\ref{I000}), as a function of
decompressed trap frequency $\omega_0$. The remaining parameter $|\beta|^{1/2}
T = 10^{-2}$, corresponds to a cycle time of 10 ms and a residual frequency of
0.3 Hz. In this case, the decompressed trap frequency is roughly half the evaporative trap frequency.

For small values of $\omega_0 \ll 1 $, the echo strength vanishes because the
weak trap creates a large cloud, which experiences more de-phasing due to the
residual potential. On the other hand, when $\omega_0 \gg 1$, the echo strength vanishes
because the tight trap increases the temperature of the cloud, resulting in a shorter echo duration.

\begin{figure}[H]
\centering
\includegraphics[scale=.7]{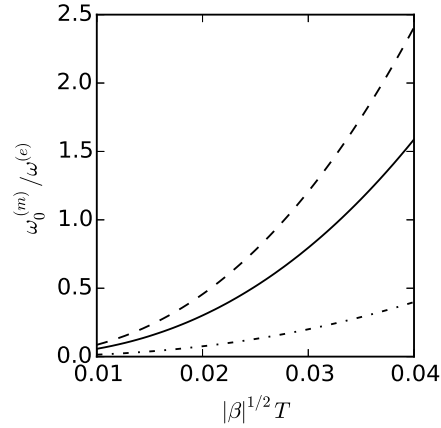}
\caption{
The ratio of the ideal injection trap frequency, $\omega_0$, to the evaporation trap frequency, $\omega^{(e)}$, as a function of $|\beta|^{1/2} T$. Here we use $\omega^{(e)} = 2\pi \times 10^{-3}$, and plot for temperatures $\mathcal{T} = 1$ (dash-dot), $\mathcal{T} = 10$ (solid), and $\mathcal{T} = 20$ (dash).  As the ratio $\omega_0/\omega^{(e)}$ becomes greater than one, the gas should be compressed before being released into the interferometer.  This compression step raises the temperature, but reduces the size of the cloud.
}
\label{fig:IofBT2}
\end{figure}

The ratio of ideal starting trap frequency $\omega_0$ and evaporation trap frequency
$\omega^{(e)}=2 \pi \times 10^{-3}$ is shown as  as a function of $|\beta|^{1/2} T$
in Figure~\ref{fig:IofBT2}. The dash-dot line is the ideal
frequency if the gas is cooled to a temperature of $\mathcal{T} = 1$, the solid line is the ideal frequency when  $\mathcal{T}=10$, and the dashed line is when $\mathcal{T} = 20$.

For values where $\omega_0 / \omega_e < 1$, the ideal starting frequency is
lower than the evaporation frequency, \emph{i.e.},\ the gas should be decompressed
before the beginning of the interferometer cycle. At the cost of increasing the cloud size, it is more advantageous the
lower the temperature. For the case where $\omega_0 / \omega_e > 1$, the gas should be compressed, raising the
temperature by reducing the size of the cloud.

\section{Outlook}

\begin{figure}[H]
\centering
\includegraphics[scale=.7]{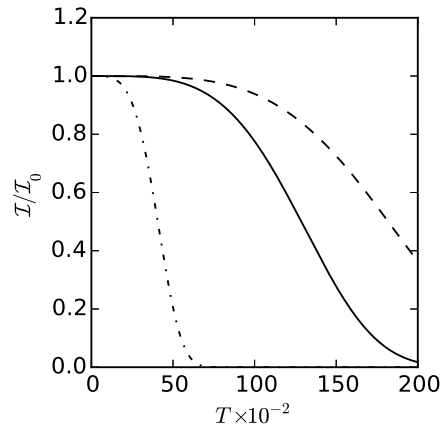}
\caption{
    The signal visibility, \emph{i.e.}, the decaying envelope that limits the maximum possible signal strength for a given interferometer time $T$.  The decay is proportional to $\mbox{exp}(-\beta^2)$, causing rapid signal loss for even small residual potentials.  Here we show $\beta =  5 \times 10^{-13}$ (dashed), $10^{-12}$ (solid), and $10^{-11}$ (dash-dot). $T \times 10^{-2}$ corresponds roughly to 1 ms.
}
\label{fig:IoT}
\end{figure}

Tuning the interferometer time $T$ and the injection trap frequency $\omega_0$ allows for maximal signal visibility. However, these optimizations cannot overcome the $\mbox{exp}(-\beta^2)$ dependence in Equation (\ref{I0}). Even a small residual potential dramatically reduces coherence times in this version of a trapped Talbot-Lau interferometer. Figure \ref{fig:IoT} shows the signal visibility, Equation (\ref{I0}), for several residual potentials. The dashed curve is $\beta = 5 \times 10^{-13}$, the solid line is $10^{-12}$, and the dash-dot curve is $10^{-11}$. For this plot, $\mathcal{I}_0 = \pi^{1/2}\mathcal{A}^2/32\mathcal{T}^{3/2}$, with $\mathcal{A} = 1/2$ to correspond to maxima in the signal oscillation. For the time axis, $T \times 10^{-2}$ corresponds roughly to 1 ms.

Clearly the signal visibility has a strong dependence on residual potential, which must be extremely small for coherence times compared to free space interferometers. In future work, we will explore modifications to the trapped Talbot-Lau scheme with the potential to minimize the coherence time's sensitivity on residual field imperfections.

The Wigner function approach allows a straightforward way to model interference in an incoherent system such as a cold atomic gas. It can be readily applied to consider different pulse schemes such as those of \cite{Wu2007}, as well as propagation in more complex confining potentials. The Talbot-Lau interferometer's ability to operate at thermal temperatures is a significant enough benefit to a real-world device that further study is warranted.

\vspace{6pt}

\acknowledgments{This work was supported by the Air Force Research Laboratory.}

\authorcontributions{Eric Imhof, James Stickney, and Matthew Squires contributed equally to this paper.}

\conflictofinterests{The authors declare no conflict of interest.}


\bibliographystyle{mdpi}
\renewcommand\bibname{References}



\end{document}